\documentclass[twocolumn,showpacs,preprintnumbers,amsmath,amssymb]{revtex4}

\usepackage{graphicx}
\usepackage{dcolumn}
\usepackage{bm}

\begin{document}

\preprint{APS/123-QED}

\title{Bose Glass-BEC Transition of Magnons in Tl$_{1-x}$K$_x$CuCl$_3$}

\author{Fumiko Yamada$^1$}
\author{Hidekazu Tanaka$^1$} 
\author{Toshio Ono$^1$}
\author{Hiroyuki Nojiri$^2$}

\affiliation{
$^1$Department of Physics, Tokyo Institute of Technology, Meguro-ku, Tokyo 152-8551, Japan\\
$^2$Institute for Material Research, Tohoku University, Aoba-ku, Sendai 980-8577, Japan
}

\date{\today}

\begin{abstract}
We report the magnetic-field induced Bose glass$-$BEC transition of magnons in Tl$_{1-x}$K$_x$CuCl$_3$ and its critical behavior investigated through specific heat and ESR measurements. The field dependence of the BEC transition temperature $T_{\rm N}$ can be described by the power law $\left[H\,{-}\,H_{\rm c}\right]\,{\propto}\,T_{\rm N}^{\phi}$ near the quantum critical point $H_{\rm c}\,{\sim}\,3.5$ T. The critical exponent ${\phi}$ tends to reach a value smaller than 1/2 with decreasing fitting window in contrast with ${\phi}\rightarrow 3/2$ for the standard BEC in pure system. At sufficiently low temperatures, the ESR line shape for $H\,{\simeq}\,H_{\rm c}$ is intermediate between Gaussian and Lorentzian. This indicates the localization of magnons for $H\,{<}H_{\rm c}$ at $T=0$. 
\end{abstract}

\maketitle


Heisenberg antiferromagnets composed of spin dimers often have gapped ground states at zero magnetic field and undergo quantum phase transitions to ordered states in magnetic fields \cite{Rice,Giamarchi}. The field-induced magnetic ordering is typical of the quantum phase transition and has been actively studied both theoretically and experimentally \cite{O_mag,Nikuni,Rueegg,O_k,Jaime,Matsumoto2,Kawashima,Misguich,Stone,Yamada,DellfAmore,Dodds}. These studies have established that the coupled antiferromagnetic spin-dimer system in a magnetic field can be mapped onto a system of interacting lattice bosons, where the bosons are spin triplets with $S_z\,{=}\,1$ called magnons or triplons, and that the field-induced magnetic ordering is best described as the Bose-Einstein condensation (BEC) of magnons. However, for the case with exchange disorder, the nature of the ground state in a magnetic field and the critical behavior of the field-induced magnetic ordering are not sufficiently understood, although there have been some studies on this problem \cite{O_T,Shindo,Nohadani_bg,Roscilde2,T_Goto,Suzuki,Hong}.

In a magnetic field $H$ comparable to the gap, the effective Hamiltonian of magnons is expressed as
\begin{eqnarray}
{\cal H}&=&\sum_i \left(J_i - g{\mu}_{\rm B}H \right)a_i^{\dagger} a_i \nonumber\\
&+& \sum_i \sum_j t_{ij} a_i^{\dagger} a_j + \frac{1}{2} \sum_i \sum_j U_{ij} a_i^{\dagger} a_j^{\dagger} a_j a_i.
\label{eq:model}
\end{eqnarray}
The first, second and third terms denote the local potential, hopping and interaction of magnons, respectively. The intradimer exchange interaction $J_i$ on dimer site $i$ corresponds to the local potential of magnons, $V_i\,{=}\,J_i\,{-}\,g{\mu}_{\rm B}H$. The ground state and the quantum phase transition for lattice bosons in a random potential were investigated theoretically by Fisher {\it et al.}~\cite{Fisher}. They argued that a new phase called Bose glass (BG) emerges as a ground state in addition to BEC and Mott insulating (MI) phases, which correspond to the ordered phase and gapped phase in the magnetic system, respectively. Bosons are localized in the BG phase because of randomness, but there is no gap; thus the compressibility is finite. Fisher {\it et al.} showed that the BEC transition occurs only from the BG phase, and that near $T\,{=}\,0$, the relation between transition temperature $T_{\rm c}$ and boson density ${\rho}$ is expressed as 
$
T_{\rm c}\,{\sim}\,[{\rho}_{\rm s}(0)]^y,\ {\rho}_{\rm s}(0)\,{\sim}\,({\rho}\,{-}\,{\rho}_{\rm c})^{\zeta},
$
where $\rho_{\rm c}$ is the critical density at which the BEC transition occurs and $\rho_{\rm s}(0)$ is the condensate density at $T\,{=}\,0$. Exponents $y$ and $\zeta$ are given by $y\,{=}\,3/4$ and ${\zeta}\,{\geq}\, 8/3$ for three dimensions. The critical behavior is different from that of the standard BEC case without randomness, for which these exponents are $y\,{=}\,2/3$ and ${\zeta}\,{=}\,1$. 

\begin{figure*}[t]
\includegraphics[width=17cm, clip]{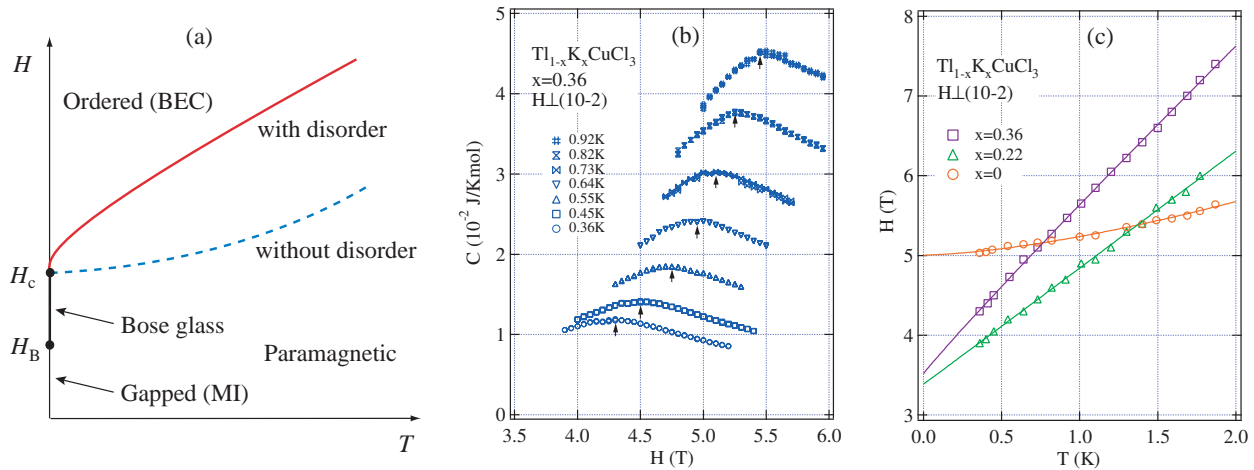}
\caption{(a) Schematic low-temperature phase boundaries expected for Tl$_{1-x}$K$_x$CuCl$_3$. Solid and dashed lines denote the boundaries for $x\neq 0$ and $x=0$, respectively. For simplification, the critical fields $H_{\rm c}$ for both the cases are plotted at the same position. (b) Field scans of the specific heat in Tl$_{1-x}$K$_x$CuCl$_3$ with $x\,{=}\,0.36$ measured at various temperatures. Arrows denote the transition field $H_{\rm N}(T)$. (c) Magnetic field vs temperature diagram obtained below 2 K for $x\,{=}0$, 0.22 and 0.36. Solid lines denote the fits by eq.\,(\ref{eq:powerlaw}) for $T\,{<}\,2$ K.}
 \label{fig:phase}
\end{figure*}

Recent theory has demonstrated the emergence of the BG phase in a disordered quantum magnet \cite{Nohadani_bg,Roscilde2}. The BG phase has also been studied in other disordered quantum systems, such as $^4$He adsorbed on porous Vycor glass \cite{Fisher,Crooker}, amorphous superconductors \cite{Okuma} and trapped cold atoms \cite{Fallani}.
From the correspondence between the boson density and the magnetization in the coupled spin-dimer system, a relation $({\rho}\,{-}\,{\rho}_{\rm c})\,{\propto}\,(H\,{-}\,H_{\rm c})$ is obtained, where $H_{\rm c}$ is the critical magnetic field of the magnon BEC. Hence, for the coupled spin-dimer system with exchange disorder, the transition field $H_{\rm N}(T)$ near $T\,{=}\,0$ should be expressed by the power law
\begin{eqnarray}
\left[H_{\rm N}(T)-H_{\rm c}\right]\,{\propto}\,T^{\phi}, 
\label{eq:powerlaw}
\end{eqnarray}
with the critical exponent ${\phi}\,{\leq}\,1/2$. Consequently, the low-temperature phase boundary should be tangential to the field axis at $T\,{=}\,0$, as shown in Fig.~\ref{fig:phase}(a). This phase boundary behavior is qualitatively different from that of pure system, for which ${\phi}\,{=}\,3/2$, and thus, the phase boundary is perpendicular to the field axis. $H_{\rm B}$ in Fig.~\ref{fig:phase}(a) is the critical field for the MI-BG transition. The BG phase exists between $H_{\rm B}$ and $H_{\rm c}$.

To investigate the magnon BEC under the influence of the localization, we performed specific heat and ESR measurements on Tl$_{1-x}$K$_x$CuCl$_3$. The parent compounds TlCuCl$_3$ and KCuCl$_3$ have the same crystal structure composed of the chemical dimer Cu$_2$Cl$_6$, in which Cu$^{2+}$ ions have spin-1/2. Their magnetic ground states are spin singlets with excitation gaps $\Delta /k_{\rm B}$ of 7.5 K and 31 K, respectively \cite{Shiramura}. The gaps originate from the strong antiferromagnetic exchange interaction between spins in the chemical dimer. The neighboring spin dimers couple antiferromagnetically in three dimensions. The intradimer exchange interaction was evaluated to be $J/k_{\rm B}\,{=}\,65.9$ K and 50.4 K for TlCuCl$_3$ and KCuCl$_3$, respectively, \cite{Cavadini1,Cavadini2,O_inela}. Because these two intradimer interactions are different, the partial K$^+$ ion substitution for Tl$^+$ ions produces the random local potential $V_i$ of the magnon \cite{note}.

Single crystals of Tl$_{1-x}$K$_x$CuCl$_3$ were synthesized from a melt comprising a mixture of TlCuCl$_3$ and KCuCl$_3$ in the ratio of $1\,{-}\,x$ to $x$. The potassium concentration $x$ was determined by inductively coupled plasma optical emission spectrochemical analysis at the Center for Advanced Materials Analysis, Tokyo Institute of Technology. 
The specific heat was measured down to 0.36 K in magnetic fields up to 9 T using a physical property measurement system (Quantum Design PPMS) by the relaxation method. 
High-frequency, high-field ESR measurements were performed using the terahertz electron spin resonance apparatus (TESRA-IMR) \cite{Nojiri} at the Institute for Material Research, Tohoku University. The temperature of the sample was lowered to 0.55 K using liquid ${}^3$He. A magnetic field was applied using a multilayer pulse magnet. In all the present experiments, the magnetic field was applied perpendicular to the cleavage (1,\,0,\,$\bar{2}$) plane, because the magnetic anisotropy in this plane is so small that the $U(1)$ symmetry is approximately conserved \cite{Glazkov}.

The specific heat obtained from the temperature scan exhibits a small cusplike anomaly due to magnetic ordering for $H\,{\geq}\,4.5$ T. The anomaly becomes smaller with decreasing magnetic field. Thus, we performed the field scan of the specific heat at various temperatures below 2 K. Some examples of the measurements for $x\,{=}\,0.36$ are shown in Fig.~\ref{fig:phase}(b). The specific heat exhibits a cusplike anomaly, to which we assign the transition field $H_{\rm N}(T)$. The transition field is well-defined within an error ${\pm}\,0.1$ T. The phase transition points obtained below 2 K for $x\,{=}0$, 0.22 and 0.36 are summarized in Fig.\,\ref{fig:phase}(c). It is clear that the critical behaviors of the phase boundaries for $x\,{=}\,0$ and $x\,{\neq}\,0$ near $T\,{=}\,0$ are qualitatively different. The phase boundary for $x\,{=}\,0$ is normal to the field axis for $T\,{\rightarrow}\,0$ but is not for $x\,{\neq}\,0$.

\begin{figure}[t]
\includegraphics[width=7.0cm, clip]{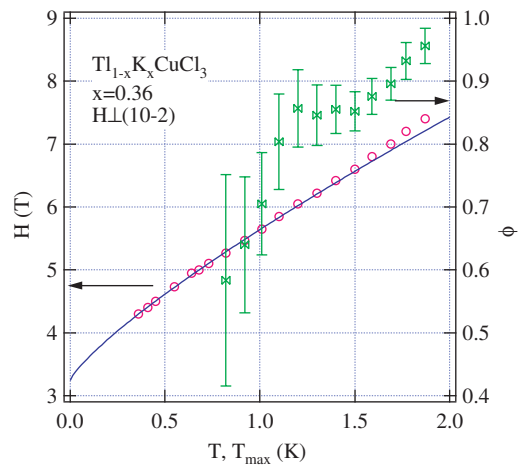}
\caption{Low-temperature phase boundary for $x\,{=}0.36$ and critical exponent $\phi$ as a function of $T_{\rm max}$ obtained by fitting eq.\,(\ref{eq:powerlaw}) to transition points between 0.36 K and $T_{\rm max}$ for $H\,{\perp}\,(1,\,0,\,\bar 2)$. The solid line is the fit with $\phi\,{=}\,0.80$.}
 \label{fig:phi036}
\end{figure}

The phase boundaries shown in Fig.\,\ref{fig:phase}(c) can be described by the power law of eq.~(\ref{eq:powerlaw}). For $x\,{=}\,0$, we obtain ${\phi}\,{=}\,1.53\,{\pm}\,0.14$ using the data for $0.35\,{\leq}\,T\,{<}\,2$ K. The solid line for $x\,{=}\,0$ in Fig.\,\ref{fig:phase}(c) is the fit with ${\phi}\,{=}\,1.53$. This critical exponent ${\phi}$ coincides with $\phi_{\rm BEC} =3/2$ predicted by the BEC theory without disorder \cite{Nikuni,Kawashima,Misguich,Fisher}. On the other hand, for $x\,{=}\,0.22$ and 0.36, we obtain ${\phi\,{=}\,}1.00\,{\pm}\,0.06$ and $0.96\,{\pm}\,0.03$, respectively, using all the data below 2 K. The solid lines for $x\,{=}\,0.22$ and 0.36 in Fig.\,\ref{fig:phase}(c) are the fits with these exponents. However, for $x\,{\neq}\,0$, the critical exponent ${\phi}$ tends to decrease with decreasing fitting window. 
Thus, we analyze the critical behavior of the phase boundary for $x\,{=}\,0.36$ in detail. We fit eq.\,(\ref{eq:powerlaw}) in the temperature range of $T_{\rm min}\,{\leq}\,T\,{\leq}\,T_{\rm max}$, setting the lowest temperature at $T_{\rm min}\,{=}\,0.36$ K and varying the highest temperature $T_{\rm max}$ from 1.87 K to 0.82 K. The critical exponent $\phi$ as a function of $T_{\rm max}$ is shown in Fig.\,\ref{fig:phi036}. The critical exponent $\phi$ decreases systematically with decreasing $T_{\rm max}$, and $\phi\,{=}\,0.58\,{\pm}\,0.17$ for $T_{\rm max}\,{=}\,0.82$ K. The solid line in Fig.\,\ref{fig:phi036} is the fit with $\phi\,{=}\,0.80\,{\pm}\,0.07$ obtained for $T_{\rm max}\,{=}\,1.1$ K. With decreasing $T_{\rm max}$ used for fitting, the critical exponent ${\phi}$ shows a clear tendency to reach a value smaller than 1/2. This critical behavior is consistent with that for the BG-BEC transition discussed by Fisher {\it et al.}~\cite{Fisher}. The similar behavior for the critical exponent is also observed for $x\,{=}\,0.22$.

As shown in Fig.\,\ref{fig:phase}(c), the critical field $H_{\rm c}$ for $x\,{=}\,0.22$ and 0.36 is smaller than that for $x\,{=}\,0$. For the following reason, this should be mainly attributed to the fact that the average intradimer interaction decreases with increasing $x$. Within the framework of the dimer mean-field approximation\,\cite{O_k}, the triplet gap is expressed as ${\Delta}\,{=}\,[J(J - 2|{\tilde J}|)]^{1/2}$, where ${\tilde J}$ is expressed by a linear combination of interdimer interactions \cite{O_k}. The gap shrinks either when the intradimer interaction $J$ is reduced or when the interdimer interaction is enhanced. The temperature $T_{\rm{M}}$ giving the maximum magnetic susceptibility decreases with increasing $x$, as shown in Fig.\,1 in Ref.~\cite{O_T}. This indicates that the average intradimer interaction decreases with $x$, because $T_{\rm{M}}$ is given by $1.60\,T_{\rm{M}}\,{=}\,J/k_{\rm B}$ in the mean-field approximation. 

\begin{figure*}[t]
\includegraphics[width=17cm, clip]{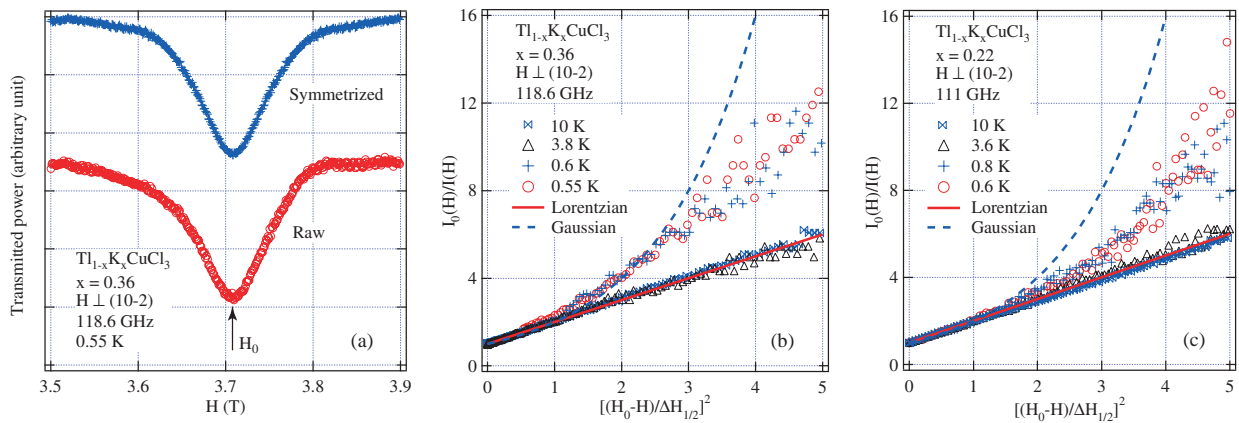}
\caption{(a) ESR spectra for $x\,{=}\,0.36$ observed at 0.55 K for 118.6 GHz. Lower and upper spectra are raw and symmetrized ones, respectively. $I(H_0)/I(H)$ vs $[(H\,{-}\,H_0)/(\Delta H_{1/2})]^2$ plots for the ESR spectra measured at various temperatures (b) for $x\,{=}\,0.36$ and (c) for $x\,{=}\,0.22$. Solid and dashed lines are the same plots for Lorentzian and Gaussian functions.}
 \label{fig:ESR}
\end{figure*}

To investigate the localization of magnons, we performed ESR measurements on Tl$_{1-x}$K$_x$CuCl$_3$ with $x\,{=}\,0.22$ and 0.36. The crystals used in the specific heat and ESR measurements were taken from the same batch. We used thick samples to gain intensities. The measurements were performed mainly at frequencies of 111 and 118.6 GHz for $x\,{=}\,0.22$ and 0.36, respectively. The paramagnetic resonance fields for these frequencies are estimated to be 3.55 and 3.80 T, using $g\,{=}\,2.23$ for $H\,{\perp}\,(1,\,0,\,{\bar 2})$. The resonance field $H_0$ is close to the critical field $H_{\rm c}$. A single ESR line was observed in the paramagnetic phase for $0.55\,{\leq}\,T\,{\leq}\,16$ K. Although the signal intensity decreases monotonically down to 0.55 K, the intensity is sufficient, even at the lowest temperature, to analyze the lineshape as shown in Fig.\,\ref{fig:ESR}. The temperature dependence of the ESR intensity is similar to that of the magnetization \cite{O_T}. We also measured ESR spectrum at 102 GHz, but we could not obtain well-defined signal to analyze the lineshape.

The lineshape of the ESR spectrum is a typical subject in the statistical mechanics of the irreversible process\,\cite{Anderson,Kubo}. In the paramagnetic phase, the linewidth is produced by the local field due to the perturbation that do not commute with the total spin such as the anisotropic exchange, dipolar interaction and the Zeeman interaction with nonuniform $g$ factors. From a statistical point of view, it is natural to assume that the local field has a Gaussian distribution. Thus, when the magnons created on the dimers localize, the lineshape should be Gaussian, because the distribution of the local field determines the lineshape of the ESR spectrum. On the other hand, if the magnons delocalize in the crystal, the local field acting on spins is rapidly averaged, so that the ESR spectrum is narrowed and becomes Lorentzian \cite{Anderson,Kubo}. This is called exchange narrowing. 

Raw ESR spectrum is somewhat unsymmetrical with respect to the resonance field $H_0$ as shown in Fig.\,\ref{fig:ESR}(a). This should be ascribed to the interference of the submillimeter wave inside the sample, because the wavelength of the submillimeter wave used and the sample thickness are the same order. Since the present system is insulator, this unsymmetrical lineshape is not intrinsic to sample. Thus, we symmetrized the spectrum, averaging both sides of $H_0$, i.e., we put $\{I(H)\,{+}\,I(2H_0\,{-}\,H)\}/2$ as the intensity $I(H)$ at $H$. The symmetrized spectrum has the flat baseline.
To analyze the lineshape of these ESR spectra, we plot $I(H_0)/I(H)$ against $[(H\,{-}\,H_0)/(\Delta H_{1/2})]^2$, where $\Delta H_{1/2}$ is the line width at $I(H_0)/2$ \cite{Dietz}. When the lineshape is Gaussian, $I(H_0)/I(H)$ increases exponentially, while for the Lorentzian lineshape, $I(H_0)/I(H)$ is a linear function of $[(H\,{-}\,H_0)/(\Delta H_{1/2})]^2$ with a slope of unity,  as shown by dashed and solid lines, respectively, in Fig.\,\ref{fig:ESR}(b) and (c). The $I(H_0)/I(H)$ vs $[(H\,{-}\,H_0)/(\Delta H_{1/2})]^2$ plots at various temperatures for $x\,{=}\,0.36$ and 0.22 are shown in Fig.\,\ref{fig:ESR}(b) and (c), respectively. We can see the similar temperature variation in these plots. 
$I(H_0)/I(H)$ at high temperatures is almost linear in $[(H\,{-}\,H_0)/(\Delta H_{1/2})]^2$ with a slope of unity, which shows the Lorentzian lineshape. On the other hand, at sufficiently low temperatures, $I(H_0)/I(H)$ vs $[(H\,{-}\,H_0)/(\Delta H_{1/2})]^2$ plot is between those of Lorentzian and Gaussian. Such an ESR lineshape is observed in one-dimensional Heisenberg antiferromagnet, in which the time correlation function of the local fields acting on the spins does not damp rapidly but has the long time tail \cite{Dietz}.

In the present temperature range, the BEC phase is not reached even at the lowest temperature. At high temperatures, thermal hopping of magnons is activated. This causes the rapid averaging of the local fields, which leads to the narrowed Lorentzian lineshape. At sufficiently low temperatures, the thermal effect is suppressed. Thus, the intermediate lineshape observed for $T\,{\leq}\,0.8$ K can be attributed to the localization of magnons. Since the resonance fields for 111 and 118.6 GHz are close to the critical field $H_{\rm c}$, it is considered that the complete localization does not occur in the present temperature range, so that the lineshape is intermediate between Gaussian and Lorentian. 

Oosawa and Tanaka \cite{O_T} reported the results of magnetization measurements on Tl$_{1-x}$K$_x$CuCl$_3$ with $x\,{\leq}\,0.36$. The magnetization curve that they observed at $T=1.8$ K had a finite slope even for $H\,{<}\,H_{\rm c}$. This was not ascribed to the finite-temperature effect, because pure TlCuCl$_3$ exhibits almost zero magnetization up to the critical field $H_{\rm c}$. This result indicates that the magnetic susceptibility ${\chi}\,{=}\,\partial M/\partial H$ for $H\,{<}\,H_{\rm c}$ is finite in the ground state. Since the magnetic susceptibility corresponds to the compressibility of the lattice boson system ${\kappa}\,{=}\,\partial \rho/\partial \mu$, where $\mu$ is the chemical potential, the finite magnetic susceptibility for $T\,{\rightarrow}\,0$ means that the compressibility of the ground state is finite. In the low-field phase below $H_{\rm c}$, long-range magnetic ordering is absent in spite of the finite susceptibility. 
These properties for $H\,{<}\,H_{\rm c}$ are consistent with the characteristics of the BG phase \cite{Fisher}. From these observations and the present ESR results, we can deduce that the ground state for $H\,{<}\,H_{\rm c}$ in Tl$_{1-x}$K$_x$CuCl$_3$ is the BG phase of magnons. The gapped MI phase for $H\,{<}\,H_{\rm B}$ appears to be destroyed in the present system.

In conclusion, from the analysis of the phase transition data and the lineshape of ESR spectrum in Tl$_{1-x}$K$_x$CuCl$_3$ combined with the previous result for magnetization measurement, we demonstrated that the quantum phase transition at $H_{\rm c}$ is the BG-BEC transition of magnons and that the critical behavior for the temperature dependence of the transition field near $H_{\rm c}$ is described by the small exponent characteristic of the BG-BEC transition \cite{Fisher}. 

We express our sincere thanks to M. Oshikawa, T. Suzuki, A. Oosawa, and T. Goto for discussions and comments. This work was supported by a Grant-in-Aid for Scientific Research (A) from the Japan Society for the Promotion of Science (JSPS), and a Global COE Program ``Nanoscience and Quantum Physics'' at TIT funded by the Ministry of Education, Culture, Sports, Science and Technology of Japan. T.O. was supported by a Grant-in-Aid for Young Scientists (B) from JSPS.


\begin{thebibliography}{99}
\bibitem{note} This partial ion substitution produces not the binomial distribution for $V_i$ but a continuous distribution, because the configuration of K$^+$ and Tl$^+$ ions around the chemical dimer Cu$_2$Cl$_6$ determines the intradimer interaction.
\end{thebibliography}
\end{document}